\documentclass[pdflatex,sn-mathphys-num]{sn-jnl}

\usepackage{graphicx}%
\usepackage{hyperref}
\usepackage{multirow}%
\usepackage{amsmath,amssymb,amsfonts}%
\usepackage{amsthm}%
\usepackage{mathrsfs}%
\usepackage[title]{appendix}%
\usepackage{xcolor}%
\usepackage{textcomp}%
\usepackage{manyfoot}%
\usepackage{booktabs}%
\usepackage{algorithm}%
\usepackage{algorithmicx}%
\usepackage{algpseudocode}%
\usepackage{listings}%
\usepackage{lineno}
\usepackage{placeins}

\theoremstyle{thmstyleone}%
\theoremstyle{thmstyletwo}%

\theoremstyle{thmstylethree}%

\begin{document}

\title[JWST-IMFERNO]{Hidden mass in early galaxies revealed by bottom-heavy initial mass functions} 

\author*[1]{\fnm{Chloe M.} \sur{Cheng}}\email{cheng@strw.leidenuniv.nl}

\author[1]{\fnm{Martje} \sur{Slob}}

\author[1]{\fnm{Mariska} \sur{Kriek}}

\author[2]{\fnm{Aliza G.} \sur{Beverage}}

\author[3]{\fnm{Pieter G.} \sur{van Dokkum}}

\author[4]{\fnm{Rachel} \sur{Bezanson}}

\author[5, 6]{\fnm{Gabriel} \sur{Brammer}}

\author[7]{\fnm{Charlie} \sur{Conroy}}

\author[7, 8]{\fnm{Anna} \sur{de Graaff}}

\author[1]{\fnm{Elham} \sur{Eftekhari}}

\author[9]{\fnm{Robert} \sur{Feldmann}}

\author[1]{\fnm{Wout M.} \sur{Goesaert}}

\author[10, 11]{\fnm{Meng} \sur{Gu}}

\author[12, 13, 14]{\fnm{Joel} \sur{Leja}}

\author[15]{\fnm{Brian} \sur{Lorenz}}

\author[1]{\fnm{Pavel E.} \sur{Mancera Pi\~na}}

\author[16, 17]{\fnm{Ignacio} \sur{Mart\'in-Navarro}}

\author[2]{\fnm{Andrew B.} \sur{Newman}}

\author[18]{\fnm{Sedona H.} \sur{Price}}

\author[19]{\fnm{Alice E.} \sur{Shapley}}

\author[1]{\fnm{Piyush} \sur{Sharda}}

\author[20]{\fnm{Katherine A.} \sur{Suess}}

\author[21]{\fnm{Arjen} \sur{van der Wel}}

\author[15]{\fnm{Daniel R.} \sur{Weisz}}

\affil*[1]{Leiden Observatory, Leiden University, \orgaddress{\street{P.O. Box 9513}, \postcode{2300RA} \city{Leiden}, \country{The Netherlands}}}

\affil[2]{\orgdiv{Observatories of the Carnegie Institution for Science}, \orgaddress{\street{813 Santa Barbara Street}, \city{Pasadena}, \state{CA} \postcode{91101}, \country{USA}}}

\affil[3]{\orgdiv{Astronomy Department}, \orgname{Yale University}, \orgaddress{\street{52 Hillhouse Ave}, \city{New Haven}, \state{CT} \postcode{06511}, \country{USA}}}

\affil[4]{\orgdiv{Department of Physics \& Astronomy and PITT PACC}, \orgname{University of Pittsburgh}, \orgaddress{\city{Pittsburgh}, \state{PA} \postcode{15260}, \country{USA}}}

\affil[5]{\orgdiv{Cosmic Dawn Center (DAWN)}, \city{Copenhagen}, \country{Denmark}}

\affil[6]{\orgdiv{Niels Bohr Institute}, \orgname{University of Copenhagen}, \orgaddress{\street{Jagtvej 128}, \postcode{2200} \city{Copenhagen N}, \country{Denmark}}}

\affil[7]{\orgdiv{Center for Astrophysics $|$ Harvard \& Smithsonian}, \orgaddress{\city{Cambridge}, \state{MA} \postcode{02138}, \country{USA}}}

\affil[8]{\orgdiv{Max-Planck-Institut f\"ur Astronomie}, \orgaddress{\street{K\"onigstuhl 17}, \postcode{D-69117} \city{Heidelberg}, \country{Germany}}}

\affil[9]{\orgdiv{Department of Astrophysics}, \orgname{Universität Zürich}, \orgaddress{\city{Zurich} \postcode{CH-8057}, \country{Switzerland}}}

\affil[10]{\orgdiv{Department of Astronomy}, \orgname{Tsinghua University}, \orgaddress{\city{Beijing} \postcode{100084}, \country{People's Republic of China}}}

\affil[11]{\orgdiv{Hong Kong Institute for Astronomy \& Astrophysics}, \orgaddress{\street{Pokfulam Road}, \orgname{The University of Hong Kong}, \city{Pok Fu Lam}, \country{Hong Kong}}}

\affil[12]{\orgdiv{Department of Astronomy \& Astrophysics}, \orgname{The Pennsylvania State University}, \orgaddress{\city{University Park}, \state{PA} \postcode{16802}, \country{USA}}}

\affil[13]{\orgdiv{Institute for Computational \& Data Sciences}, \orgname{The Pennsylvania State University}, \orgaddress{\city{University Park}, \state{PA} \postcode{16802}, \country{USA}}}

\affil[14]{\orgdiv{Institute for Gravitation and the Cosmos}, \orgname{The Pennsylvania State University},  \orgaddress{\city{University Park}, \state{PA} \postcode{16802}, \country{USA}}}

\affil[15]{\orgdiv{Department of Astronomy}, \orgname{University of California}, \orgaddress{\city{Berkeley}, \state{CA} \postcode{94720}, \country{USA}}}

\affil[16]{\orgdiv{Instituto de Astrofísica de Canarias}, \orgaddress{\street{c/ Vía Láctea s/n}, \postcode{E38205} - \city{La Laguna}, \state{Tenerife}, \country{Spain}}}

\affil[17]{\orgdiv{Departamento de Astrofísica}, \orgname{Universidad de La Laguna}, \orgaddress{\postcode{E-38205} \city{La Laguna}, \state{Tenerife}, \country{Spain}}}

\affil[18]{\orgname{Space Telescope Science Institute}, \orgaddress{\street{3700 San Martin Drive}, \city{Baltimore}, \state{MD} \postcode{21218}, \country{USA}}}

\affil[19]{\orgdiv{Department of Physics \& Astronomy}, \orgname{University of California, Los Angeles}, \orgaddress{\street{430 Portola Plaza}, \city{Los Angeles}, \state{CA} \postcode{90095}, \country{USA}}}

\affil[20]{\orgdiv{Department for Astrophysical \& Planetary Science}, \orgname{University of Colorado, Boulder}, \orgaddress{\state{CO} \postcode{80309}, \country{USA}}}
\affil[21]{\orgname{Sterrenkundig Observatorium}, \orgname{Universiteit Ghent}, \orgaddress{\street{Krijgslaan 281 S9}, \postcode{B-9000} \city{Gent}, \country{Belgium}}}

\abstract{James Webb Space Telescope (JWST) observations have revealed that massive galaxies formed and evolved faster than predicted by galaxy formation models, with many having already assembled a large mass in stars approximately 12 billion years ago.  However, masses of distant galaxies are uncertain, as they assume a distribution of stellar birth masses (the initial mass function (IMF)) similar to that in the Milky Way.  Specifically, the contribution from low-mass stars, which make up the bulk of stellar mass, is not directly observed, but inferred based on an extrapolation of the Milky Way IMF.  Here, we provide robust constraints on the low-mass IMF beyond the local Universe from full-spectrum models.  Using ultra-deep spectra of nine massive quiescent galaxies at redshift $z$ $\approx0.7$ from the JWST Initial Mass Function of Early Red NIRSpec Objects program, extended to bluer wavelengths with deep Very Large Telescope Large Early Galaxy Astrophysics Census spectra, we find that the most massive galaxies have excess low-mass stars.  Remarkably, our oldest galaxy (formation redshift $z_{\rm form} > 5$) has the most bottom-heavy IMF.  This galaxy may be a descendant of JWST's `impossibly early' galaxies, implying that the latter may have had similarly bottom-heavy IMFs increasing their masses by a factor of approximately $4\pm1$.  Our findings may thus amplify the tension with galaxy formation models.}

\maketitle

The IMF is an essential ingredient in galaxy formation models.  It is required to infer fundamental properties from observed galaxy light, including stellar masses and star formation rates.  Historically, the IMF has been considered to be `universal', and assumed to take the same form as the Milky Way (MW) IMF everywhere.  However, in agreement with theoretical predictions (see, for example, refs. \cite{Schwarzschild_1953, Kroupa, Krumholz_2011, Hopkins_2012, Chabrier_2014}), observational evidence that the IMF varies in diverse stellar systems in the nearby Universe (see, for example, refs. \cite{Spiniello_2012, Cappellari_2013, Geha_2013, La_Barbera_2013, McDermid_2014, Martin-Navarro_2015, Martin-Navarro_2021, Hallakoun_2021, Cheng_2023} and also ref. \cite{Smith_2020}) has been mounting.  In particular, the cores of present-day massive early-type galaxies have been found to host bottom-heavy IMFs (see, for example, refs. \cite{CvD12b, van_dokkum_2017, Gu_2022}).  Despite these indications, studies of distant massive quiescent galaxies -- the likely progenitors of nearby early-type cores -- still largely adopt a single, MW IMF.  Thus, the assumed shape of the IMF represents a major source of systematic uncertainty in our mass estimates.  With recent James Webb Space Telescope (JWST) observations revealing significant tensions between the inferred stellar masses of early galaxies and galaxy formation models (see, for example, refs. \citep{Steinhardt_2016, Carnall_2023a, Carnall_2023b, Carnall_2024, Antwi-Danso_2025, Baker_2025, de_graaff_2025, Ito_2025, Stevenson_2026}), it is imperative that we constrain the IMF at early times.

\begin{figure}[h!]
    \centering
    \includegraphics[width=\linewidth]{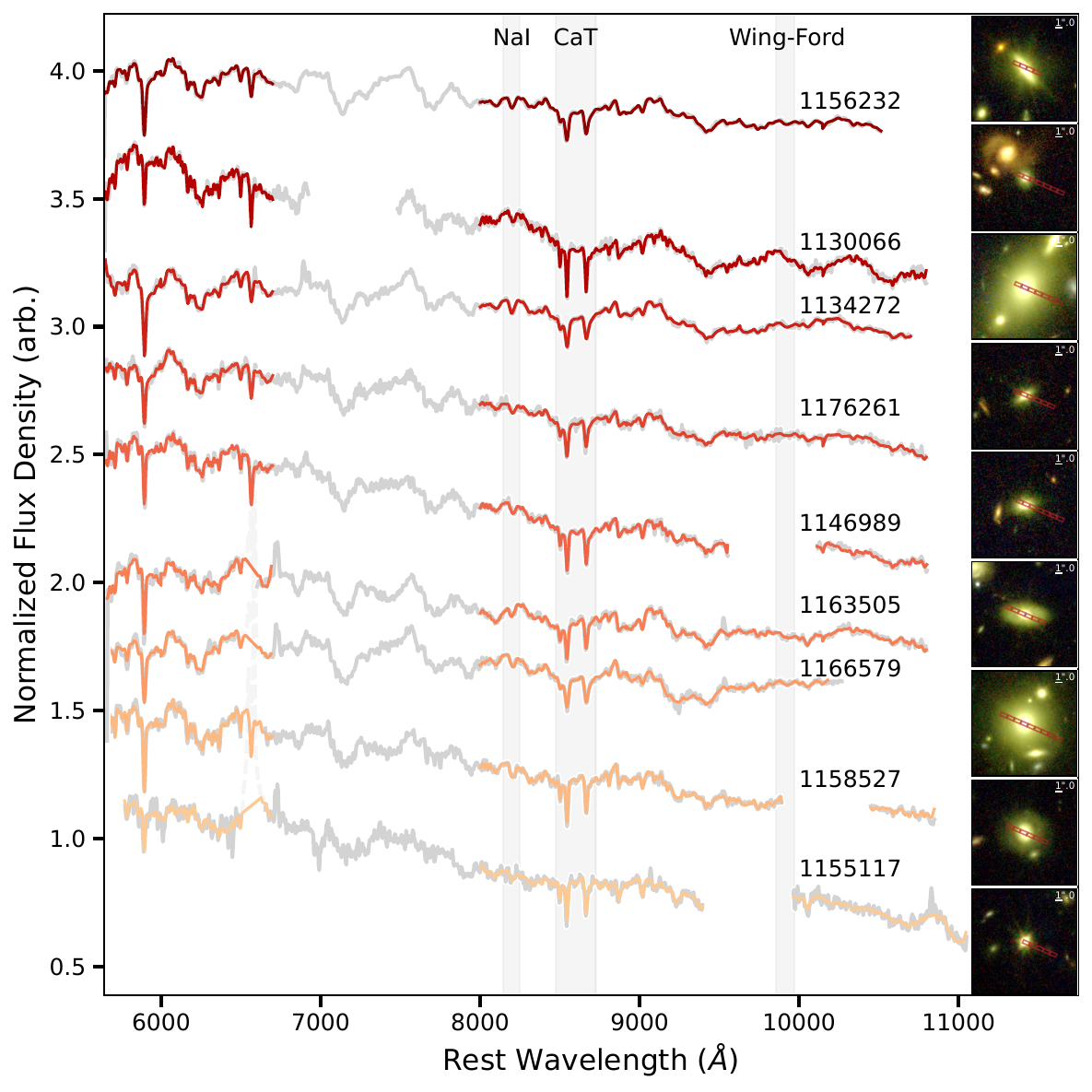}
    \caption{\textbf{JWST-NIRSpec/MSA spectra with best-fit stellar population models and image cut-outs.}  The NIRSpec/MSA spectra with best-fit stellar population models over-plotted are shown for our sample of 9 massive quiescent galaxies at $z\approx0.7$, observed via the IMFERNO program.  Spectra are presented in order of increasing redshift (bottom to top).  We normalize each spectrum by the median flux between rest-frame $7,500$ and $7,600$ \AA, arbitrarily offsetting them in the $y$ direction for visibility.  We indicate key IMF-sensitive absorption features with shaded regions.  We do not fit the region between $7,000$ and $8,000$ \AA\ owing to broad TiO absorption (see Methods).  We also show the red-green-blue (RGB) image cutout ($8^{\prime\prime}.5 \times 8^{\prime\prime}.5$) for each galaxy next to its corresponding spectrum, where we have combined the F115W, F277W, and F444W COSMOS-Web images \citep{cosmos-web, grizli, Valentino_2023}.  We display the NIRSpec/MSA shutters on top of each image.}
    \label{fig:spectra}
\end{figure}

In this Article, we constrain the low-mass end of the stellar IMF in nine massive quiescent galaxies at $z\approx0.7$, observed via the JWST Initial Mass Function of Early Red NIRSpec Objects (IMFERNO) program (PID 5629).  Targets were selected from the Large Early Galaxy Astrophysics Census (LEGA-C) \citep{van_der_Wel_2021}, on the basis of their quiescent stellar populations, redshifts, $H$-band magnitudes, and availability of Cosmic Evolution Survey (COSMOS)-Web photometry (refs. \citep{cosmos-web, grizli, Valentino_2023} and Methods). The IMFERNO spectra were obtained with the JWST Near Infrared Spectrograph micro shutter assembly (NIRSpec/MSA), using the G140M-F100LP disperser-filter combination.  Combined with the bluer LEGA-C spectra, the average rest-frame wavelength coverage is approximately $3,700 - 10,800$ \AA.  To robustly constrain the low-mass end of the IMF, we require a median signal-to-noise ratio (S/N) $\gtrsim15$ \AA$^{-1}$ for the LEGA-C spectra and a median S/N $\gtrsim70$ \AA$^{-1}$ for the IMFERNO spectra, which we determine via mock spectra tests similar to ref. \cite{Cheng_2024}.  On the LEGA-C side, this rest-frame S/N allows us to constrain ages and elemental abundance patterns required to break strong degeneracies between abundance and IMF variations \citep{CvD_2012a, CvD12b, Gu_2022}.  On the IMFERNO side, we require higher rest-frame S/N as we target much fainter spectral features, sensitive to the relative fraction of red dwarfs to red giants.  We show the nine IMFERNO spectra in Fig.~\ref{fig:spectra}, and the corresponding LEGA-C spectra in Extended Data Fig.~\ref{fig:legac_fits}.  We show an example combined LEGA-C + IMFERNO spectrum in Extended Data Fig.~\ref{fig:fit_figure}.  Our sample properties are reported in Extended Data Table~\ref{tab:sample}.  

To constrain the low-mass end of the IMF, we simultaneously fit the entire rest-frame wavelength range covered by LEGA-C and IMFERNO with the \textsc{absorption line fitter} (\textsc{alf}), a flexible, full-spectrum stellar population synthesis (SPS) code \citep{CvD_2012a, Conroy_2018}.  \textsc{alf} is unique among SPS codes in that it can account for variability in abundance patterns and the low-mass shape of the IMF.  This flexibility, along with our broad wavelength range, allows us to disentangle degeneracies between elemental abundance and low-mass IMF variations \citep{CvD_2012a, CvD12b, Gu_2022}.  We adopt a single-age stellar population, where age, 18 elemental abundances, and the two low-mass slopes of a double broken power-law IMF can vary freely (Methods and Extended Data Fig.~\ref{fig:imf_slopes}).  We show representative corner plots of key recovered parameters for two galaxies in our sample in Extended Data Fig.~\ref{fig:corner_plots}.  We find that we are able to disentangle well-known degeneracies between the low-mass IMF and elemental abundance variations (refs. \cite{CvD_2012a, CvD12b, Gu_2022} and Methods).  To enable comparison with previous work, we also fit the spectra with a Kroupa IMF \cite{Kroupa}, referred to as the MW IMF going forward.  We test adopting a two-component star-formation history (SFH) composed of two bursts of star formation with free ages and a parameter representing the relative mass fraction of the young component.  The results from this more complex SFH are consistent within uncertainties with those of the single-age model.  Thus, we present only the single-age model.  

We show the fits to the IMFERNO spectra in Fig.~\ref{fig:spectra}, along with galaxy red-green-blue (RGB) images from COSMOS-Web \citep{cosmos-web, grizli, Valentino_2023}.  We present the LEGA-C fits in Extended Data Fig.~\ref{fig:legac_fits}, although note that we fit the LEGA-C and IMFERNO spectra for each galaxy simultaneously.  We also show an example full spectrum and best-fitting model in Extended Data Fig.~\ref{fig:fit_figure}.

\begin{figure}[h!]
    \centering
    \includegraphics[width=\linewidth]{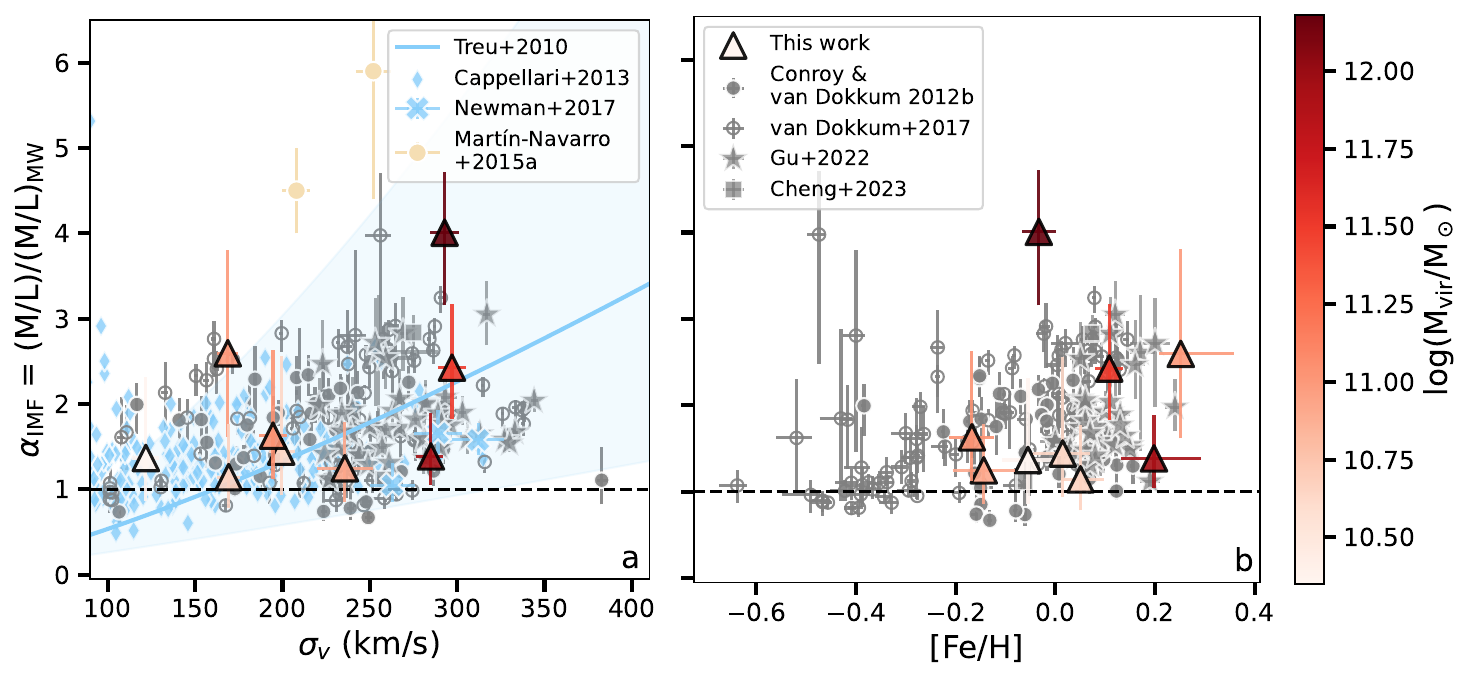}
    \caption{\textbf{The IMF mismatch parameter compared to different galaxy properties.}  \textbf{a, b} The IMF mismatch parameter ($\alpha_{\rm IMF}$), defined as the ratio between the $M/L$ ratio where we allow the low-mass IMF to vary freely and the $M/L$ ratio where we assume a MW IMF, shown as a function of the stellar velocity dispersion ($\sigma_v$; \textbf{a}) and iron abundance ([Fe/H]; \textbf{b}).  The IMFERNO galaxies are shown as triangles, colour-coded by their virial masses ($\log({\rm M}_{\rm vir}/{\rm M}_{\odot})$).  Small open and closed circles, stars, and squares represent measurements for massive elliptical galaxies in the nearby Universe, from ref. \cite{CvD12b} (re-derived in ref. \cite{Cheng_2023}), ref. \cite{van_dokkum_2017}, ref. \cite{Gu_2022}, and ref. \cite{Cheng_2023}.  In \textbf{a}, the small diamonds, cross, and solid line represent lensing or dynamical measurements in the nearby Universe, from refs. \citep{Treu_2010, Cappellari_2013, Newman_2017}.  Large circles in \textbf{a} represent spectral index constraints from stacked spectra at $z\approx1.1$, from ref. \citep{Martin_Navarro_2015_highzimf}.  Error bars and uncertainty bands represent statistical uncertainties at the level of one standard deviation.}
    \label{fig:alpha_2panel}
\end{figure}

Our \textsc{alf} fit results are summarized in Fig.~\ref{fig:alpha_2panel} and reported in Extended Data Table~\ref{tab:results}.  We show the IMF mismatch parameter ($\alpha_{\rm IMF}$), which is the ratio between the mass-to-light ($M/L$) ratio where we allow the low-mass end of the IMF to vary freely and the $M/L$ ratio where we adopt a MW IMF.  As the luminosity is the same for each IMF characterization, $\alpha_{\rm IMF}$ represents the stellar mass excess compared to the MW.  We derive $M/L$ ratios in the rest-frame Hubble Space Telescope/ACS-F814W band.  The dashed line represents where galaxies with a MW IMF would lie, with anything above this line considered to be bottom-heavy.  We colour-code each point by virial mass ($\log({\rm M}_{\rm vir}/{\rm M}_{\odot})$; see ref. \citep{van_der_wel_2022} and Methods).  We compare $\alpha_{\rm IMF}$ values to integrated stellar velocity dispersions ($\sigma_v$; Methods) and [Fe/H] ratios (from our \textsc{alf} fits).  For comparison, we show analogous literature measurements from nearby massive elliptical galaxies (in grey \citep{CvD12b, van_dokkum_2017, Gu_2022, Cheng_2023}) and lensing/dynamical measurements from nearby massive ellipticals (in blue \citep{Treu_2010, Cappellari_2013, Newman_2017}).  Figure~\ref{fig:alpha_2panel} demonstrates that two of our galaxies have $\alpha_{\rm IMF}$ values that are $> 2$ standard deviations above 1, indicating that they have an excess of low-mass stars compared to the MW (Extended Data Fig.~\ref{fig:imf_slopes}).  In other words, their IMFs are bottom-heavy.  When considering the full sample, we find that $\alpha_{\rm IMF}$ ranges from 1.1 to 4.0, with a median value of 1.4.  We primarily find bottom-heavy IMFs in galaxies with higher $\sigma_v$, [Fe/H], and $\log({\rm M}_{\rm vir}/{\rm M}_{\odot})$.  Qualitatively similar trends were found in the nearby Universe, using similar and complementary methods (see refs. \citep{Treu_2010, CvD12b, Cappellari_2013, La_Barbera_2013, McDermid_2014, van_dokkum_2017} and also ref. \citep{Smith_2020}).  Thus, we show that these trends were already in place at $z\approx0.7$, 6.5 billion years ago.  Our results are consistent with weak lensing \cite{Fujikawa_2026} and spectroscopic indications \cite{Martin_Navarro_2015_highzimf} beyond the nearby Universe.  In particular, ref. \cite{Martin_Navarro_2015_highzimf} constrained the IMF in two galaxy stacks at $z\approx1.1$ using the IMF-sensitive TiO$_2$ index (Fig.~\ref{fig:alpha_2panel}, beige circles).  While our individual-galaxy constraints from full-spectrum fitting are more robust, it is encouraging that distant galaxy stacks exhibit qualitatively similar IMF behaviour.

\begin{figure}
    \centering
    \includegraphics[width=0.7\linewidth]{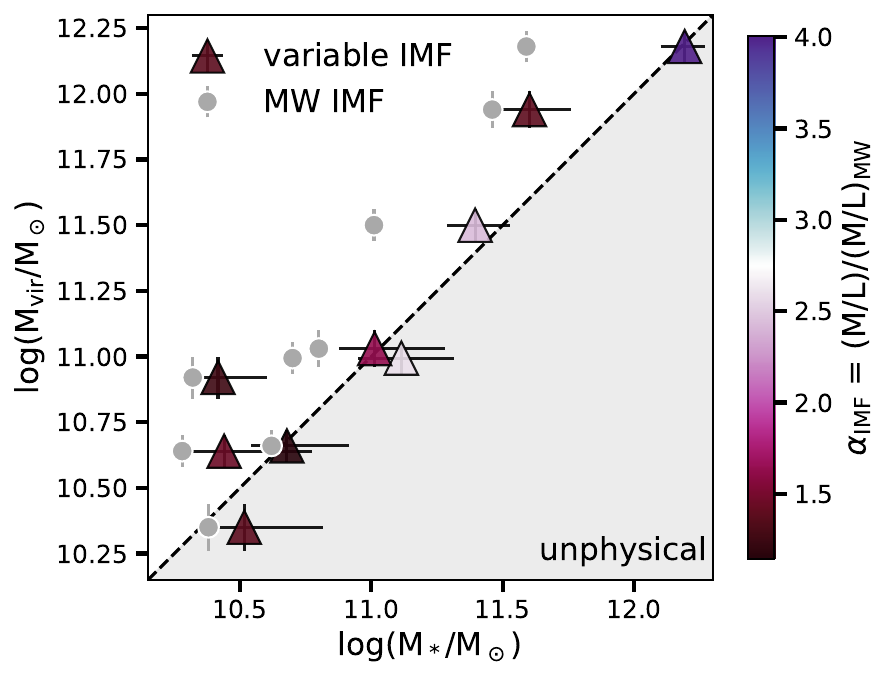}
    \caption{\textbf{Comparison between stellar and virial masses for different low-mass IMF parameterizations.}  Virial masses ($\log({\rm M}_{\rm vir}/{\rm M}_{\odot})$) are shown on the $y$ axis (Methods).  Stellar masses, derived using \textsc{MAGPHYS} (\citep{MAGPHYS, de_graaff_2021}) and assuming a MW IMF, are shown on the $x$ axis (circles).  We correct these masses for our best-fit low-mass IMF by multiplying them by $\alpha_{\rm IMF}$.  We show the corrected masses as triangles, colour-coded by $\alpha_{\rm IMF}$.  Error bars represent statistical uncertainties at the level of one standard deviation.  The shaded region represents an unphysical region, where stellar masses are larger than virial masses.  The dashed line represents a one-to-one comparison.}
    \label{fig:mass_compare}
\end{figure}

In our analysis we only studied the stellar components of our galaxies, and it is important to consider whether our inferred IMFs are realistic when accounting for non-baryonic matter.  Thus, to assess the plausibility of our findings, we compare virial (Methods) and stellar masses in Fig.~\ref{fig:mass_compare}.  We show stellar masses, estimated using \textsc{MAGPHYS} \citep{MAGPHYS} and assuming a MW IMF \citep{de_graaff_2021}, as grey circles.  We correct the \textsc{MAGPHYS} masses for our best-fit IMF by multiplying them by $\alpha_{\rm IMF}$, and show these corrected masses as triangles colour-coded by $\alpha_{\rm IMF}$.  Figure~\ref{fig:mass_compare} shows that, when assuming a MW IMF, the offset between virial and stellar masses increases towards higher mass.  Similar results were found in refs. \citep{van_de_sande_2013, Belli_2017, Mendel_2020, Kriek_2024, Slob_2025}.  Interestingly, the stellar masses increase when applying our variable IMF, with the galaxies with the largest virial-to-stellar mass offsets shifting the most.  This stellar mass increase generally brings the stellar and virial masses into agreement.  In particular, the stellar masses do \textit{not} exceed the virial masses within uncertainties, and are consistent with the physical plausibility of bottom-heavy IMFs in these galaxies (see also ref. \citep{Lyubenova_2016}).

The excellent consistency between the virial and IMF-corrected stellar masses in Fig.~\ref{fig:mass_compare} leaves little room for dark matter and gas.  However, this is not unexpected as these galaxies have small effective radii (a median of approximately $5.4$ kpc) and thus the virial masses primarily capture their star-dominated central regions (see, for example, ref. \citep{de_graaff_2024}).  Additionally, little interstellar gas is expected to be present in massive quiescent galaxies at these redshifts (see, for example, ref. \citep{Spilker_2018}).  Moreover, the virial masses may be underestimated as we do not fully correct $\sigma_v$ for slit-loss effects (Methods and ref. \citep{Slob_2025}).  In addition, these galaxies may be rotating \citep{Bezanson_2018, van_Houdt_2021}, in which case their $\sigma_v$ values may not be an ideal tracer of their potential wells, contributing to the possible underestimation of the virial masses.  This underestimation would allow for more non-baryonic matter.  Detailed dynamical modelling is required to fully account for these effects (ref. \citep{Slob_2025}, and Slob et al. in preparation).

Our finding of bottom-heavy IMFs at early times is consistent with the two-phase formation model for massive galaxies, in which the cores of massive quiescent galaxies are in place at early times (with their central densities staying constant), with their outskirts subsequently building up via minor mergers \citep{Bezanson_2009, Naab_2009, Feldmann_2010, Oser_2010, Rodriguez_Gomez_2016, Suess_2020}.  Moreover, a few of our galaxies may be even more bottom-heavy than nearby galaxy cores.  Thus, in the two-phase scenario, the mixing of stellar populations by merging activity may reduce the bottom-heavy nature of the IMF between $z\approx0.7$ and $z\approx0$ in such galaxies (see also ref. \citep{Chabrier_2014}).

We examine the galaxies for which we find the most bottom-heavy IMFs (that is, inconsistent with a MW IMF by $> 2$ standard deviations based on statistical uncertainties) in detail.  Galaxy 1134272 has the oldest age and earliest formation time of our sample, implying a formation redshift of $z_{\rm form} \gtrsim 5$.  In general, we would expect older galaxies to have slightly higher $\alpha_{\rm IMF}$ values, as an excess of low-mass stars fractionally contributes more to the mass for an older stellar population (in which more of the high-mass stars have died).  However, this effect would be $< 10$\% for the ages covered by our sample and thus cannot explain the high $\alpha_{\rm IMF}$.  Assuming a typical star formation timescale of approximately $100-200$ million years \citep{Kriek_2016, Beverage_suspense}, this oldest galaxy was quiescent by $z\approx4.6$, and thus may be a descendant of the `impossibly early' galaxies discovered with JWST \citep{Carnall_2023a, Carnall_2023b, Carnall_2024, Valentino_2023, de_graaff_2025, Ito_2025}.  We note that we do not exclude the possibility that the remaining galaxies in our sample are also descendants of these early, massive galaxies, as their relatively younger ages could be driven by late-time accretion or star formation.  In any case, the formation redshift of 1134272 as the oldest and most bottom-heavy galaxy implies that JWST's distant, massive galaxies may have also had bottom-heavy IMFs at early times.  If similarly bottom-heavy IMFs were already in place when these galaxies formed, this would make them a factor of approximately $4\pm1$ more massive (considering statistical uncertainties alone) than originally reported.  These larger masses would suggest an even greater tension with galaxy formation models (\citep{Steinhardt_2016, Carnall_2023a, Carnall_2023b, Carnall_2024, Valentino_2023, Antwi-Danso_2025, de_graaff_2025, Ito_2025}). 

In this context, we note that we have applied a single $\alpha_{\rm IMF}$ to each galaxy, implicitly assuming that the low-mass IMFs (and other stellar population parameters) measured within the NIRSpec/MSA slitlets are representative of each galaxy as a whole.  However, present-day massive early-type galaxies -- the likely descendants of our distant systems -- are known to exhibit IMF gradients, with bottom-heavy cores (within $0.4$ effective radii ($R_{\rm e}$)) and MW-like outskirts (see, for example, refs. \citep{CvD12b, van_dokkum_2017, Gu_2022} but also ref. \citep{Alton_2017}).  On average, our spectra probe the inner half-light radius of our galaxies (Extended Data Fig.~\ref{fig:fit_figure}), meaning that the outer regions of our galaxies covered by our slitlets could plausibly host a different IMF.  In a scenario where the low-mass IMF becomes fully MW-like beyond $1{\ }R_{\rm e}$, our correction factors should actually be $\alpha_{\rm IMF}/2$.  However, given the substantial size growth observed over cosmic time (a factor of $\gtrsim2$ in half-light radius between $z\approx5$ and $z\approx0.7$; see, for example, refs. \citep{Bezanson_2009, Valentino_2023, Baker_2025, Stevenson_2026}) and assuming limited stellar mixing, the central regions of our galaxies were likely already in place at higher redshifts.  Under this assumption, applying a uniform $\alpha_{\rm IMF}$ to the full galaxies at $z$ of approximately $4-5$ would be justified.  Nevertheless, our low-mass IMF constraints here should be taken to be aperture-specific.  Spatially resolved IMF measurements in the IMFERNO galaxies will be the subject of future work. 

In this Article, we have presented novel measurements of the low-mass end of the IMF in the early Universe, using state-of-the-art JWST data and sophisticated full-spectrum modelling methods.  However, some limitations should be considered when interpreting our results.  First, we note that our sample size is small, and that our galaxies are associated with an over-density (Methods).  These factors may limit the generalizability of our findings.  In the future, larger and more diverse sample sizes are required to confirm our findings.

Additionally, our findings are largely based on our \textsc{alf} fits to the IMFERNO spectra.  While \textsc{alf} is currently the only code capable of constraining the shape of the low-mass IMF from continuum-normalized spectra while also accounting for variable abundance patterns, there may be some level of systematic uncertainty as a result of our modelling choices (i.e. stellar evolution assumptions, model and prior choices within the available parameter space, continuum-normalisation, line-spread function modelling; see, for example, refs. \citep{Conroy_2013, Byrne_2023, Bellstedt_2025}).  In particular, different spectral libraries predict different absorption feature strengths in the rest-frame optical at the same age and metallicity \citep{Byrne_2023, Bellstedt_2025, Jones_2025}.  Encouragingly, ref. \citep{Byrne_2023} found the C3K stellar library (used by \textsc{alf}) to be the most robust out of the libraries that they tested.  Additionally, previous work has shown that Lick index and full-spectrum modelling methods generally produce the same results in local galaxies within uncertainties (see, for example, refs. \citep{Villaume_2017b, Lonoce_2021, Cheng_2023}).  We have also performed several sensitivity tests (Methods), and find that our constraints are generally robust to our precise fitting setup (that is, IMF low-mass cutoff, masking of IMF sensitive features, and continuum normalization).  Furthermore, we note that our comparison between virial and stellar masses in Fig.~\ref{fig:mass_compare} supports the plausibility of our results.  

We also fix the slope of the IMF above $1{\ }{\rm M}_\odot$, as our method is only sensitive to the relative fraction of low-to-high-mass stars \citep{CvD_2012a}.  However, if this slope is shallower than in the MW, the stellar masses at high redshifts can be reduced.  We emphasize that our finding of bottom-heavy IMFs does not negate the possibility that the slope of the IMF beyond $1{\ }{\rm M}_\odot$ also differed from that of the MW at formation.  In particular, the IMF across different stellar mass regimes may respond differently to the physical conditions of star formation, which could affect inferred stellar masses and their implications for galaxy formation models (see, for example, refs. \cite{Larson_1998, Fontanot_2018, La_Barbera_2019, Martin-Navarro_2021}).  For example, previous studies have proposed IMFs that are simultaneously top- and bottom-heavy (see, for example, refs. \cite{Arrigoni_2010, Coulter_2017, den_brok_2024, skislope}), time-varying (see, for example, refs. \cite{Vazdekis_1997, Martin-Navarro_2016, Jerabkova_2018}), or metallicity-dependent (see, for example, refs. \cite{Kroupa, Yan_2024}). 

Though our spectra have an unprecedented depth at these redshifts and our IMF measurements are supported by the comparison with virial masses in Fig.~\ref{fig:mass_compare}, our results are still uncertain.  In an upcoming study, we will perform detailed anisotropic Jeans modelling of the IMFERNO galaxies to independently probe our conclusions (Slob et al. in preparation). Additionally, by extending our spectra to longer wavelengths and examining redder IMF-sensitive spectral features, the IMF can be more precisely constrained.  Redder IMF constraints will also require further development of stellar population models, as they are not currently able to fully reproduce such spectra of nearby galaxies \citep{Eftekhari_2021, Eftekhari_2022}.  This spectral extension and development of the near-infrared models will be the subject of an upcoming approved JWST Cycle 5 program to obtain redder spectra of the current IMFERNO sample (PID 10660).

In summary, we robustly measured the low-mass end of the IMF in a sample of massive quiescent galaxies beyond the nearby Universe using full-spectrum modelling of extremely deep JWST/NIRSpec-MSA data.  Our finding of bottom-heavy IMFs at $z\approx0.7$ within the apertures probed by the spectra, particularly in our oldest galaxy, suggests that the most massive quiescent galaxies discovered at high redshifts (stellar masses $\gtrsim10^{11}{\ }{\rm M}_\odot$; refs. \citep{Steinhardt_2016, Carnall_2023a, Carnall_2023b, Carnall_2024, de_graaff_2025, Ito_2025}) may have even higher masses than originally reported.  Our results could additionally imply that the masses of their star-forming progenitors may also be underestimated, further heightening the tension with galaxy formation models.  Looking forwards, IMF measurements at higher redshifts, even closer to the epoch when these `impossibly early' galaxies formed, will clarify this apparent tension with galaxy formation models.  Future extremely deep JWST/NIRSpec observations are required to make these constraints possible. 


\backmatter
\bmhead{Data Availability}
This work makes use of data from the LEGA-C survey. The reduced data can be obtained from the ESO Science Archive Facility (\url{http://archive.eso.org/eso/eso_archive_main.html}).  The reduced spectra and catalogue have been released by ESO (\url{http://archive.eso.org/cms/eso-archive-news/Third-and-final-release-of-the-Large-Early-Galaxy\\-Census-LEGA-C-Spectroscopic-Public-Survey-published.html}) and are also available via the University of Gent at \url{https://users.ugent.be/~avdrwel/research.html\#legac} (see ref. \citep{van_der_Wel_2021} for details).  The object IDs for the LEGA-C data can be found in Extended Data Table~\ref{tab:sample} (column 2), and can be used to identify the data in the ESO archive.  The JWST data analysed in this work were obtained from the Mikulski Archive for Space Telescopes (MAST) at the Space Telescope Science Institute (STScI).  The specific observations can be accessed via MAST at \href{https://doi.org/10.17909/k0zz-f371}{10.17909/k0zz-f371} (ref. \citep{JWST_doi}).  Our sample properties and fit results are published in Extended Data Tables~\ref{tab:sample} and \ref{tab:results}.  Source files for the Supplementary Information are included in Supplementary Data 1.  Files containing the \texttt{alf.f90} configuration files for each galaxy are available in Supplementary Data 2.  Files containing the spectra as they are fit with \textsc{alf} (including the resolution curves and wavelength windows) are available in Supplementary Data 3.  Source data are provided with this paper.

\bmhead{Code Availability}
The \textsc{alf} code on which this work is based is publicly available via GitHub at \href{https://github.com/cconroy20/alf}{https://github.com/cconroy20/alf} (commit c21ccc5; see refs. \cite{CvD_2012a, Conroy_2018} for details).  The procedure used to rescale the spectra to the photometry is available via GitHub at \url{https://github.com/Mslob/JWST_IMFERNO_phot_rescaling}.      

\bmhead{Acknowledgments}
We thank the anonymous referees for taking the time to give us useful feedback that improved this manuscript.  We thank the LEGA-C team for making their dataset public.  This work is based on observations made with the NASA/ESA/CSA JWST.  The data were obtained from the Mikulski Archive for Space Telescopes (MAST) at the Space Telescope Science Institute (STScI), which is operated by the Association of Universities for Research in Astronomy, Inc., under NASA contract NAS 5-03127 for JWST.  These observations are associated with program JWST-GO-5629.  This work used the Dutch national e-infrastructure with the support of the SURF Cooperative using grants no. EINF-10017 and EINF-16504, which are financed by the Dutch Research Council (NWO). Some of the data products presented herein were retrieved from the Dawn JWST Archive (DJA). DJA is an initiative of the Cosmic Dawn Center (DAWN), which is funded by the Danish National Research Foundation under grant DNRF140. 

\bmhead{Funding Statement}
MK acknowledges funding from the NWO through the award of the Vici grant VI.C.222.047.  Support for program JWST-GO-5629 was provided by NASA through a grant from the STScI.  AGB acknowledges support from NASA through the NASA Hubble Fellowship grant \#HST-HF2-51571 awarded by the Space Telescope Science Institute, which is operated by the Association of Universities for Research in Astronomy, Inc., for NASA, under contract NAS5-26555.  AdG acknowledges support from a Clay Fellowship awarded by the Smithsonian Astrophysical Observatory.  PEMP acknowledges the support from the NWO through the Veni grant VI.Veni.222.364.  PS is supported by the Leiden University Oort Fellowship and the IAU Gruber Foundation Fellowship.

\bmhead{Author Contributions} MK, AGB, and CMC wrote the primary JWST proposal.  MK was the primary supervisor of this project and critically edited the text.  MS, CMC, MK, and AGB designed the JWST observing plan.  MS reduced the data and wrote the data reduction text.  CMC fit the spectra with \textsc{alf}, performed the velocity dispersion correction, wrote the remaining text, and created the figures.  CMC, MK, and PGvD led the interpretation.  All authors contributed to the analysis and interpretation.  

\bmhead{Competing Interests}
The authors declare no competing interests. 






\newpage
\section*{Methods}\label{sec11}

\subsection*{Observing strategy and data reduction}
To robustly constrain the low-mass IMF, we require extremely deep spectra covering rest-frame wavelengths between approximately $4,000$ and $10,000$ \AA. To achieve this beyond the local Universe, we take advantage of the optical spectra from the third data release of LEGA-C, a European Southern Observatory (ESO) Public Spectroscopic survey of 3600 galaxies at $0.6\lesssim z\lesssim 1.0$. These deep (20-h integration), $R\approx3500$ spectra were collected over 128 nights using the VIsible MultiObject Spectrograph (VIMOS) on the ESO Very Large Telescope (\textit{VLT}), resulting in an average S/N of approximately $20$ \AA$^{-1}$ (see ref. \cite{van_der_Wel_2021} for details).

At $z\approx0.7$, the LEGA-C observing strategy results in a rest-frame wavelength range of approximately $3,700 - 5,100$ \AA, capturing several Balmer features and metal lines that are sensitive to age, [Fe/H], and several elemental abundances. However, features that are sensitive to the ratio of dwarf to giant stars, which can be used to constrain the low-mass IMF \cite{CvD_2012a}, are located between approximately $8,100$ and $10,000$ \AA. To capture these features, we extend the LEGA-C spectra with extremely deep NIRSpec/MSA \citep{PFerruit2022} spectra from the JWST-IMFERNO program (PID 5629, PIs Kriek, Beverage, and Cheng), executed on 3-4 and 25 May 2025. 

To select our targets for JWST-IMFERNO, we identified quiescent targets from the LEGA-C galaxy sample with the $UVJ$ classification from ref. \cite{Muzzin_2013_uvj}, using the UltraVISTA \citep{McCracken_2012} DR3 $K_s$-band catalogue \citep{Muzzin_2013}.  In order to achieve our S/N requirements, we required an $H$-band magnitude $< 21.5$.  We additionally required the NIRSpec/MSA spectra to capture IMF sensitive features including NaI (approximately $8,180 - 8,200$ \AA), CaT (approximately $8,475 - 8,725$ \AA), and the Wing-Ford band (approximately $9,905 - 9,945$ \AA; ref. \citep{Wing_Ford_1969}) where possible.  This was feasible for galaxies at $z\approx0.7$ using the G140M-F100LP disperser-filter combination, allowing us to obtain spectra covering approximately $9,700 - 18,400$ \AA, with a spectral resolution of $R\approx1300$ (see, for example, refs. \citep{msafit,Slob_2024}).  This wavelength range corresponds to a rest-frame wavelength coverage of approximately $5,700 - 10,800$ \AA\ on average.  We identified one extraordinary pointing for which we observed 13 massive ($10.28 \lesssim \log(M_*/M_\odot) \lesssim 11.59$, assuming a Kroupa IMF \citep{Kroupa, van_der_Wel_2021}) quiescent galaxies at $z\approx0.7$ which have LEGA-C optical spectra and COSMOS-Web imaging \cite{cosmos-web}.  Nine of the 13 galaxies have sufficient S/N to simultaneously constrain ages, elemental abundances, and low-mass IMF variations.  We note that all of our targets fit in one pointing as most are part of the `COSMOS Wall' structure at $z\approx0.73$ \citep{Scoville_2007_structure}.  Thus, our targets are also located in an old region of the $z\approx0.7$ Universe, reinforcing the connection between our targets and massive galaxies in the early Universe.  

The galaxies were observed using a custom two-point nod pattern with a two-micro-shutter ($1^{\prime\prime}.06$) offset in the cross-dispersion direction, following the JWST-SUSPENSE program \cite{Slob_2024}.  With this nodding pattern and large offset, we mitigate the self-subtraction of continuum emission for our extended targets in the data reduction process.  To correct for detector and micro shutter artefacts we observed our galaxies in two nearly identical MSA configurations, offset by 3 micro shutters in the dispersion direction. Our nod-dither pattern resulted in four observing positions. While it was not possible to observe all filler targets in all four positions, seven of our nine primary targets presented in this work were covered by all positions resulting in a total on-source integration time of 31.2 h.  Two primary targets (1155117 and 1158527) were only fully observed in three nod-dither positions due to slit failures during one visit.  However, these two targets still have high enough S/N to be included in our final sample.  To ensure sufficient empty sky for background calibration, we observed the primary targets with slitlets consisting of five to nine shutters. 

We reduce the data to two-dimensional (2D) flux calibrated spectra using the JWST Science Calibration Pipeline \citep{Bushouse_2023} v1.20.2 and version 1464 of the Calibration Reference Data System. We add a 1/$f$ correlated vertical readout noise correction to this pipeline, as described in ref. \citep{Slob_2024}. After generating the 2D spectra via the pipeline, we use an optimal extraction algorithm with a Moffat profile \citep{Moffat_1969} to extract the one-dimensional (1D) spectra. Since the two dithers in our observations have slightly different MSA configurations, we reduce the dithers separately and combine the resulting 1D spectra. Finally, to ensure the LEGA-C and IMFERNO spectra are on the same flux scale, we re-scale the 1D spectra from each survey to the overlapping COSMOS2025 \citep{MShuntov2025} photometry.  

\subsection*{Full-spectrum fitting}
To measure ages, elemental abundances, and low-mass IMF slopes, we fit the combined LEGA-C and IMFERNO spectra of 9 massive, quiescent galaxies at $z\approx0.7$ with the \textsc{alf}, a full spectrum SPS model \citep{CvD_2012a, Conroy_2018}.  The \textsc{alf} models are built on empirical simple stellar populations, created using the MESA Isochrones and Stellar Tracks (MIST, \citep{MIST}) and the Spectral Polynomial Interpolator (SPI, \citep{Villaume_2017}).  We use the Medium Resolution INT Library of Empirical Spectra (MILES, \citep{Sanchez_Blazquez_2006}), the Extended Infrared Telescope Facility (E-IRTF) stellar library \citep{Villaume_2017}, and a large sample of M-dwarf spectra \citep{Mann_2015} with SPI.  We additionally use a theoretical stellar library (C3K \cite{CvD_2012a}) to ensure the quality of interpolation at the boundaries of the empirical parameter space.  The \textsc{alf} models allow for variable abundance patterns by differentially including theoretical element response functions.  We explore an empirical parameter space spanning $-2.0\lesssim {\rm [Fe/H]} \lesssim 0.5$ and $3.5\lesssim \log(T_{\rm eff}/{\rm K})\lesssim 3.9$, which is set by the combined E-IRTF and ref. \cite{Mann_2015} samples.  With these components, \textsc{alf} develops stellar spectra as a function of effective temperature ($T_{\rm eff}$), surface gravity, and metallicity from a data-driven model.  

We parameterize the IMF as a double broken power-law with break points at $m = 0.5{\ }{\rm M}_\odot$ and $1{\ }{\rm M}_\odot$, similar to the Kroupa IMF \citep{Kroupa}, and a fixed low-mass cutoff ($m_c$) at $0.08{\ }{\rm M}_\odot$.  From $1.0{\ }{\rm M}_\odot < m < 100{\ }{\rm M}_\odot$, the IMF slope is assumed to have the Salpeter \cite{Salpeter} value of 2.35:
\begin{equation}\label{eq:2PL}
    \frac{dN}{dm} = \begin{cases}
    k_1m^{-\alpha_1}, &0.08{\ }{\rm M}_\odot < m < 0.5{\ }{\rm M}_\odot \\
    k_2m^{-\alpha_2}, &0.5{\ }{\rm M}_\odot < m < 1.0{\ }{\rm M}_\odot \\
    k_3m^{-2.35}, &1.0{\ }{\rm M}_\odot \leq m < 100{\ }{\rm M}_\odot
    \end{cases}
\end{equation}
Here, $N$ is the number of stars, $m$ is stellar mass, $k_i$ are normalization constants, and $\alpha_1$ and $\alpha_2$ represent the two low-mass slopes of the IMF.  For a MW IMF, we use the Kroupa \cite{Kroupa} values of $\alpha_1 = 1.3$ and $\alpha_2 = 2.3$.  This IMF parameterization is shown in Extended Data Fig.~\ref{fig:imf_slopes}, for a Kroupa IMF, a Salpeter IMF, and the best-fit IMFs of the IMFERNO sample.  

We note that fixing the upper mass limit to $100{\ }{\rm M}_\odot$ does not substantially affect the $M/L$ ratio, especially for the old stellar populations that we are examining here.  In this case, stars near the upper mass limit have already died and evolved into remnants.  Thus, changing this limit primarily affects how much mass was lost due to stellar evolution, rather than the present-day light.  As a result, increasing the limit (for example, to $150{\ }{\rm M}_\odot$) lowers the current stellar mass by $\lesssim 2$\% while leaving the luminosity unchanged. 

Similarly, we fix the low-mass cutoff ($m_c$) to the hydrogen-burning limit at $0.08{\ }{\rm M}_\odot$.  We perform a preliminary test of the sensitivity of our results to this choice by constraining $m_c$ to a small range around $0.1{\ }{\rm M}_\odot$ and find that our results are robust to the assumed $m_c$.  However, we emphasise that we are unlikely to be able to constrain $m_c$ well, as this requires much higher S/N ($\gtrsim 300$ \AA$^{-1}$; refs. \citep{Conroy_2017, Newman_2017}).  

To derive stellar population parameters, \textsc{alf} fits a high-order Chebyshev polynomial to the ratio of the data to the model, to continuum-normalize the target spectrum.  We note that our constraints are robust to the choice of polynomial order.  We test this by fixing the order of the polynomial to $n = 7$ across the entire spectrum of each of our galaxies and refitting them.  We find that the results are consistent with our fiducial fits within uncertainties.  See refs. \cite{CvD12b, Conroy_2018} for additional \textsc{alf} continuum-normalization tests.  

We also note that NIRSpec/MSA spectra have been found to be affected by under-sampling of the NIRSpec point-spread function, resulting in broad variations in flux as a function of wavelength, known as `wiggles' \cite{Cheng_2025b}.  As we continuum-normalise our spectra, these broad features should not have a notable effect on our fitting results.  Moreover, these artefacts have only been found to notably affect spatially resolved spectra.  Nevertheless, we test correcting these wiggles in our IMFERNO spectra using the algorithm presented in ref. \cite{Cheng_2025b} and find that the wiggle correction does not substantially affect our results or change our conclusions.

Using a \textsc{fortran} implementation of the Markov chain Monte Carlo algorithm, \texttt{emcee} \citep{emcee}, \textsc{alf} samples the posteriors of 46 different stellar parameters, allowing for arbitrary variation in stellar age, elemental abundances, and the two low-mass slopes of the IMF (equation~\ref{eq:2PL}).  It also fits for systematic parameters to characterize observed errors, including a jitter term.  Additionally, it calculates $M/L$ ratios by computing the mass in stars and remnants (using the prescription in ref. \cite{Renzini_1993}), assuming an IMF, and dividing this by the integral of the spectrum over the filter of interest (Hubble Space Telescope/ACS-F814W in our case).  $M/L$ ratios are calculated assuming the best-fit free IMF and Kroupa IMF for each fit.  For details, see refs. \cite{CvD_2012a, Conroy_2018}.  On the basis of our assumptions, we note that all uncertainties stated throughout the text are statistical, and do not include systematics.

We smooth the \textsc{alf} models to the instrumental resolution of the LEGA-C and NIRSpec/MSA observations before fitting.  We determine the instrumental resolution for the bluer sides of the spectra from LEGA-C as in refs. \cite{Cheng_2024, Cheng_2025}, by modelling 40 skylines across the wavelength range of each LEGA-C spectrum as Gaussians, and converting the median Gaussian $\sigma$ to a velocity dispersion in km/s as a function of wavelength.  For the redder sides of the spectra, since the instrumental resolution of the NIRSpec/MSA depends strongly on the source morphology and position in the slit \citep{msafit, Slob_2024}, we derive the instrumental resolution for each individual galaxy using \textsc{msafit} \citep{msafit} and COSMOS-Web F150W NIRCam imaging \citep{cosmos-web}, using a double S\'ersic profile. The derived instrumental resolution for each galaxy differs from pre-launch estimates (\url{https://jwst-docs.stsci.edu/jwst-near-infrared-spectrograph/nirspec-instrumentation/nirspec-dispersers-and-filters}) for the G140M disperser, by a wavelength independent factor. We thus use the pre-launch estimated resolution curve, multiplied by the derived offset (factors ranging from 1.2 to 1.4), as the final instrumental resolution for each galaxy.  We note that we test switching the resolution curves for two of the galaxies in our sample, and find that our constraints are robust to the precise modelling of the resolution curve.

Following the methodology of refs. \cite{Cheng_2024, Cheng_2025}, we fit each spectrum using $1,024$ walkers, $20,000$ burn-in steps, and a $1,000$ step production run.  We examine $500$ Markov chain Monte Carlo chains per fit and visually confirm that they are converged.  We fit a single age, and avoid the walkers getting trapped at an unrealistically high age by initializing the age of each galaxy with a random value drawn from a uniform distribution centred at 3 Gyr.  Additionally, we set the upper limit of the age prior to be the age of the Universe at the redshift of each galaxy, plus 2 Gyr to allow for uncertainties.  We note that \textsc{alf} can fit spectra between $3,700$ and $24,000$ \AA\ and can be used for stellar populations that are older than 1 Gyr.  In Extended Data Fig.~\ref{fig:corner_plots}, we display representative corner plots for one bottom-heavy galaxy from our sample (1134272) and one Milky Way-like galaxy from our sample (1146989).  We show our constraints on age, [Z/H], [Fe/H], [Na/H], [Ca/H], and the two IMF slopes ($\alpha_1$ and $\alpha_2$) for each galaxy.

Our fiducial results consist of fits to the full spectra, fitting the LEGA-C and IMFERNO spectra simultaneously (although note that we additionally fit the LEGA-C and IMFERNO spectra separately for testing purposes; `Stellar velocity dispersions and virial masses' section).  We mask strong emission lines, including [OIII], [SII], and the H$\alpha+$[NII] complex, when they are present.  Where possible, we fit the wavelength ranges $3,700 - 4,700$ \AA, $4,700 - 5,200$ \AA, $5,600 - 6,000$ \AA, $6,000 - 6,700$ \AA, $8,000 - 8,920$ \AA, $8,920 - 9,630$ \AA, $9,630 - 10,150$ \AA, and $10,150 - 10,800$ \AA, excluding the regions between $7,000$ and $8,000$ \AA\ owing to broad TiO absorption (in line with previous work; see, for example, refs. \cite{CvD12b, Cheng_2023, Beverage_suspense, Cheng_2025b}).  We also test fitting the spectra including the region between $7,000$ and $8,000$ \AA\ and find that while the results are relatively consistent with our fiducial fit, including this TiO absorption introduces large error bars for some of our $\alpha_{\rm IMF}$ parameters.  

In our fiducial fit, we include the NaD absorption feature near $5,900$ \AA, which is important for constraining the upper limit of the IMF-degenerate Na-abundance \citep{CvD_2012a}.  However, since NaD can be affected by the interstellar medium \citep{CvD_2012a}, we also test masking this feature.  Similar to our TiO test, we find that while the results are consistent, some of the errorbars on $\alpha_{\rm IMF}$ are inflated.  We show the effect of this test on our sample in Supplementary Fig.~\ref{fig:NaD_effect}.  We note that, independent of the IMF, our constraints on [Na/H] come from a combination of the NaD and NaI features, as well as the entire spectrum.  This is due to the fact that Na has an indirect effect on the whole spectrum as a result of the free electron abundance in stellar atmospheres.  On the other hand, [Ca/H] has been found to be constrained well by the CaI feature near $4,227$ \AA.  See \cite{CvD12b} for more details.

Moreover, we note that for some of the IMFERNO spectra, the IMF-sensitive Wing-Ford band \citep{Wing_Ford_1969} falls in the detector gap (Fig.~\ref{fig:spectra}).  Nonetheless, it has been shown that while all three features (NaI, CaT, and the Wing-Ford band) are important for robustly constraining the low-mass IMF, the exact shape of the low-mass IMF and trends with bottom-heaviness are not sensitive to any one feature.  In other words, the results remain consistent when excluding individual features, as IMF variations affect the entire spectrum (see \citep{CvD12b, Cheng_2023}).  We perform similar tests (i.e. masking each of NaI, CaT, and the Wing-Ford band) and find that our constraints are similarly robust to excluding individual IMF-sensitive features.  We show our final fits to the LEGA-C spectra in Extended Data Fig.~\ref{fig:legac_fits} and our final fits to the IMFERNO spectra in Fig.~\ref{fig:spectra}, but again note that we fit the two sides of each spectrum simultaneously. 

Finally, for each galaxy we calculate $\alpha_{\rm IMF}$ (the ratio between the $M/L$ ratio where we allow the low-mass IMF to vary and the $M/L$ ratio where we fix a MW IMF) and use this $\alpha_{\rm IMF}$ value to correct the stellar mass.  These masses were originally constrained by fitting the photometric spectral energy distributions (from the UltraVISTA $K_s$-band catalogue \citep{Muzzin_2013}) of the galaxies with the \textsc{MAGPHYS} code \citep{MAGPHYS}.  They have been scaled to the total stellar mass using the total luminosity of the best-fit Sérsic profile (see \citep{de_graaff_2021}).  We correct the \textsc{MAGPHYS} masses with our best-fit $\alpha_{\rm IMF}$ values instead of extracting masses from our \textsc{alf} fits as \textsc{MAGPHYS} considers star-formation histories and dust more rigorously than \textsc{alf} (which only has two simple star-formation history options and does not account for dust). 

\subsection*{Stellar velocity dispersions and virial masses}\label{sec:kinematic_correction}
The bluer LEGA-C and redder IMFERNO data were obtained with two different instruments.  Thus, we must consider the fact that we are not examining exactly the same spatial areas of each galaxy in the blue and red regions of the spectra, due to differences in apertures, point-spread functions, slit orientations, and slit offsets with respect to galaxy centres.  In particular, our measurements may be affected by stellar population gradients \citep{Cheng_2024, Cheng_2025b, Jafariyazani_2020} and galaxy dynamics \citep{Bezanson_2018a, Slob_2025}.  Stellar population gradients in LEGA-C galaxies have been found to be relatively weak \citep{Cheng_2024}.  In contrast, the velocity dispersions may be affected by differences in slit orientation, as the LEGA-C slits are oriented in the north-south direction and the IMFERNO slitlets have a position angle of approximately $247$ degrees.  This is especially the case if the galaxies are rotating (ref. \cite{Bezanson_2018a} and Slob et al. in preparation).  

To assess -- and, if needed, correct for -- velocity dispersion differences, we fit the IMFERNO spectra in isolation (i.e. without combining them with the bluer LEGA-C spectra), and compare the recovered velocity dispersions ($\sigma_{v,\rm \ IMFERNO}$) to those from the LEGA-C survey, measured with \textsc{ppxf} (ref. \citep{Cappellari2023}, $\sigma_{v, \rm \ LEGA-C}$) \citep{van_der_Wel_2021, Bezanson_2018}.  For galaxies whose $\sigma_v$'s are offset by more than one standard deviation considering statistical uncertainties (four galaxies), we broaden the side of the spectrum (i.e. the LEGA-C or IMFERNO side) with the lower $\sigma_v$.  We convolve the respective side of the spectrum with a Gaussian kernel with a width equal to the quadrature difference between $\sigma_{v, \rm \ LEGA-C}$ and $\sigma_{v,\rm \ IMFERNO}$.  We fine-tune the precise width of the Gaussian until the two $\sigma_v$ are consistent.  We re-fit the corrected full spectra (i.e. LEGA-C + IMFERNO, shown in the body of the paper) and the individual sides of the spectra to assess the effect of our correction.  We show this effect in Supplementary Fig.~\ref{fig:sigma_correction}.  We note that this correction does not strongly affect our conclusions.

For the three galaxies for which $\sigma_{v, \rm \ IMFERNO}$ is greater than $\sigma_{v, \rm \ LEGA-C}$ (probably owing to different slit alignments along the kinematic major axes), we also correct their virial masses.  In particular, we multiply the virial masses from ref. \cite{van_der_wel_2022} by a factor of $\sigma_{v, \rm \ IMFERNO}^2/\sigma_{v, \rm \ LEGA-C}^2$.  These corrected virial masses are shown throughout the paper.  We additionally use the maximum velocity dispersion between LEGA-C and IMFERNO as the fiducial $\sigma_v$ throughout the paper, as this is more representative of the galaxy's velocity dispersion.  This is due to the fact that the lower $\sigma_v$ may neglect part of the kinematics if the slit is oriented along the minor axis.


\backmatter
\newpage
\renewcommand{\figurename}{Extended Data Figure} 
\renewcommand{\tablename}{Extended Data Table} 
\bmhead{Extended Data}
\newpage
\begin{table}[h]
\caption{\textbf{Properties of our sample of nine galaxies from JWST-IMFERNO.} \footnotetext[1]{Derived from running \textsc{SourceExtractor} \citep{SourceExtractor} on the COSMOS-Web F150W images \citep{cosmos-web}.}  \footnotetext[2]{From the LEGA-C survey \citep{van_der_Wel_2021}.}  \footnotetext[3]{From the UltraVISTA K-band photometric catalogue \citep{Muzzin_2013}.}}\label{tab:sample}%
\begin{tabular}{@{}llllll@{}}
\toprule
ID & LEGA-C ID & RA\footnotemark[1]  & Dec\footnotemark[1] & $z_{\rm spec}$\footnotemark[2] & $H$\footnotemark[3] \\
 &  & (hh:mm:ss) & (dd:mm:s) & & (mag) \\ 
\midrule
1155117	& 209642 &	09:59:40.85	&	+02:30:51.2	&	0.6984	&	20.5		 \\
1158527	& 209933	&	09:59:38.92	&	+02:30:59.7	&	0.7281	&	19.9		 \\
1166579	& 210564	&	09:59:39.77	&	+02:31:16.7	&	0.7286	&	18.6		 \\
1163505	& 239944	&	09:59:30.41	&	+02:31:07.4	&	0.7303	&	19.8		 \\
1146989	& 238668	&	09:59:33.55	&	+02:30:20.2	&	0.7324	&	20.2		 \\
1176261	& 211337	&	09:59:39.27	&	+02:32:04.2	&	0.7327	&	20.8		 \\
1134272	& 207735	&	09:59:46.85	&	+02:29:08.4	&	0.7340	&	18.4		 \\
1130066	& 207657	&	09:59:36.93	&	+02:29:20.3	&	0.7347	&	21.2		 \\
1156232	& 239469	&	09:59:32.99	&	+02:30:47.2	&	0.7357	&	19.2		 \\
\botrule
\end{tabular}
\end{table}
\newpage

\begin{sidewaystable}[h!]
\caption{\textbf{Results from our \textsc{alf} fits to the combined LEGA-C and JWST-IMFERNO spectra.}  $\sigma_v$ represents the largest $\sigma_v$ between the LEGA-C reported value and the value that we obtain from fitting the IMFERNO spectra in isolation.  $M/L$ ratios are derived in the \textit{HST}/ACS-F814W band.}\label{tab:results}%
\begin{tabular*}{\textheight}{@{\extracolsep\fill}rrrrrrrrrr}
\toprule
ID	&	$\log({\rm M}_{\rm vir})$	&	$\log({\rm M}_{*})$	&	$\log({\rm M}_{*})$\footnotemark[1]	&	$\sigma_{v}$	&	Age	&	[Fe/H]	&	$M/L$	&	$M/L$	&	$\alpha_{\rm IMF}$	 \\
	&	(M$_\odot$)	&	(free, M$_\odot$)	&	(MW, M$_\odot$)	&	(km/s)	&	(Gyr)	&		&	(free, ${\rm M}_\odot/{\rm L}_\odot$)	&	(MW, ${\rm M}_\odot/{\rm L}_\odot$)	&		 \\
\midrule																					
1155117	&	$10.64\pm0.06$	&	$10.44_{-0.15}^{+0.34}$	&	$10.28$	&	$199.0_{-7.0}^{+8.0}$	&	$1.34_{-0.07}^{+0.06}$	&	$0.01_{-0.06}^{+0.05}$	&	$9.76_{-3.41}^{+7.19}$	&	$6.73_{-0.12}^{+0.14}$	&	$1.44_{-0.51}^{+1.12}$	 \\
1158527	&	$10.99\pm0.06$	&	$11.11_{-0.16}^{+0.20}$	&	$10.70$	&	$169.0_{-6.0}^{+6.0}$	&	$2.40_{-0.07}^{+0.08}$	&	$0.25_{-0.04}^{+0.11}$	&	$28.00_{-10.67}^{+12.37}$	&	$10.65_{-0.17}^{+0.19}$	&	$2.60_{-0.98}^{+1.21}$	 \\
1166579	&	$11.94\pm0.07$	&	$11.60_{-0.11}^{+0.16}$	&	$11.46$	&	$285.0_{-8.0}^{+7.0}$	&	$3.64_{-0.20}^{+0.21}$	&	$0.20_{-0.07}^{+0.09}$	&	$19.74_{-4.38}^{+6.52}$	&	$14.35_{-0.52}^{+0.58}$	&	$1.39_{-0.34}^{+0.51}$	 \\
1163505	&	$11.03\pm0.07$	&	$11.01_{-0.14}^{+0.27}$	&	$10.80$	&	$195.0\pm8.0$\footnotemark[1]	&	$1.95_{-0.07}^{+0.07}$	&	$-0.17_{-0.05}^{+0.04}$	&	$14.79_{-4.69}^{+9.09}$	&	$8.85_{-0.15}^{+0.15}$	&	$1.63_{-0.51}^{+1.00}$	 \\
1146989	&	$10.66\pm0.06$	&	$10.68_{-0.14}^{+0.24}$	&	$10.62$	&	$169.0_{-6.0}^{+6.0}$	&	$2.10_{-0.09}^{+0.08}$	&	$0.05_{-0.04}^{+0.05}$	&	$11.00_{-3.62}^{+5.86}$	&	$9.38_{-0.15}^{+0.15}$	&	$1.14_{-0.36}^{+0.62}$	 \\
1176261	&	$10.92\pm0.08$	&	$10.42_{-0.14}^{+0.19}$	&	$10.32$	&	$236.0\pm16.0$\footnotemark[1]	&	$4.49_{-0.77}^{+1.11}$	&	$-0.14_{-0.06}^{+0.06}$	&	$17.63_{-5.25}^{+7.78}$	&	$13.99_{-1.17}^{+1.99}$	&	$1.25_{-0.39}^{+0.54}$	 \\
1134272	&	$12.18\pm0.06$	&	$12.19_{-0.09}^{+0.08}$	&	$11.59$	&	$293.0\pm8.0$\footnotemark[1]	&	$5.86_{-0.49}^{+0.51}$	&	$-0.03_{-0.03}^{+0.03}$	&	$64.56_{-13.38}^{+11.68}$	&	$16.48_{-0.76}^{+0.83}$	&	$4.01_{-0.85}^{+0.71}$	 \\
1130066	&	$10.35\pm0.09$	&	$10.52_{-0.16}^{+0.30}$	&	$10.38$	&	$122.0_{-8.0}^{+7.0}$	&	$2.95_{-0.14}^{+0.25}$	&	$-0.05_{-0.05}^{+0.05}$	&	$16.25_{-6.01}^{+11.19}$	&	$11.39_{-0.37}^{+0.48}$	&	$1.37_{-0.51}^{+0.95}$	 \\
1156232	&	$11.50\pm0.06$	&	$11.39_{-0.11}^{+0.13}$	&	$11.01$	&	$297.0\pm8.0$\footnotemark[1]	&	$3.61_{-0.22}^{+0.24}$	&	$0.11_{-0.03}^{+0.03}$	&	$31.42_{-7.83}^{+9.07}$	&	$12.93_{-0.37}^{+0.39}$	&	$2.43_{-0.60}^{+0.75}$	 \\
\botrule
\end{tabular*}
\footnotetext[1]{From the LEGA-C survey \citep{van_der_Wel_2021}, measured with \textsc{ppxf} \cite{Cappellari2023}.}
\end{sidewaystable}

\setcounter{figure}{0}
\begin{figure}
    \centering
    \includegraphics[width=\linewidth]{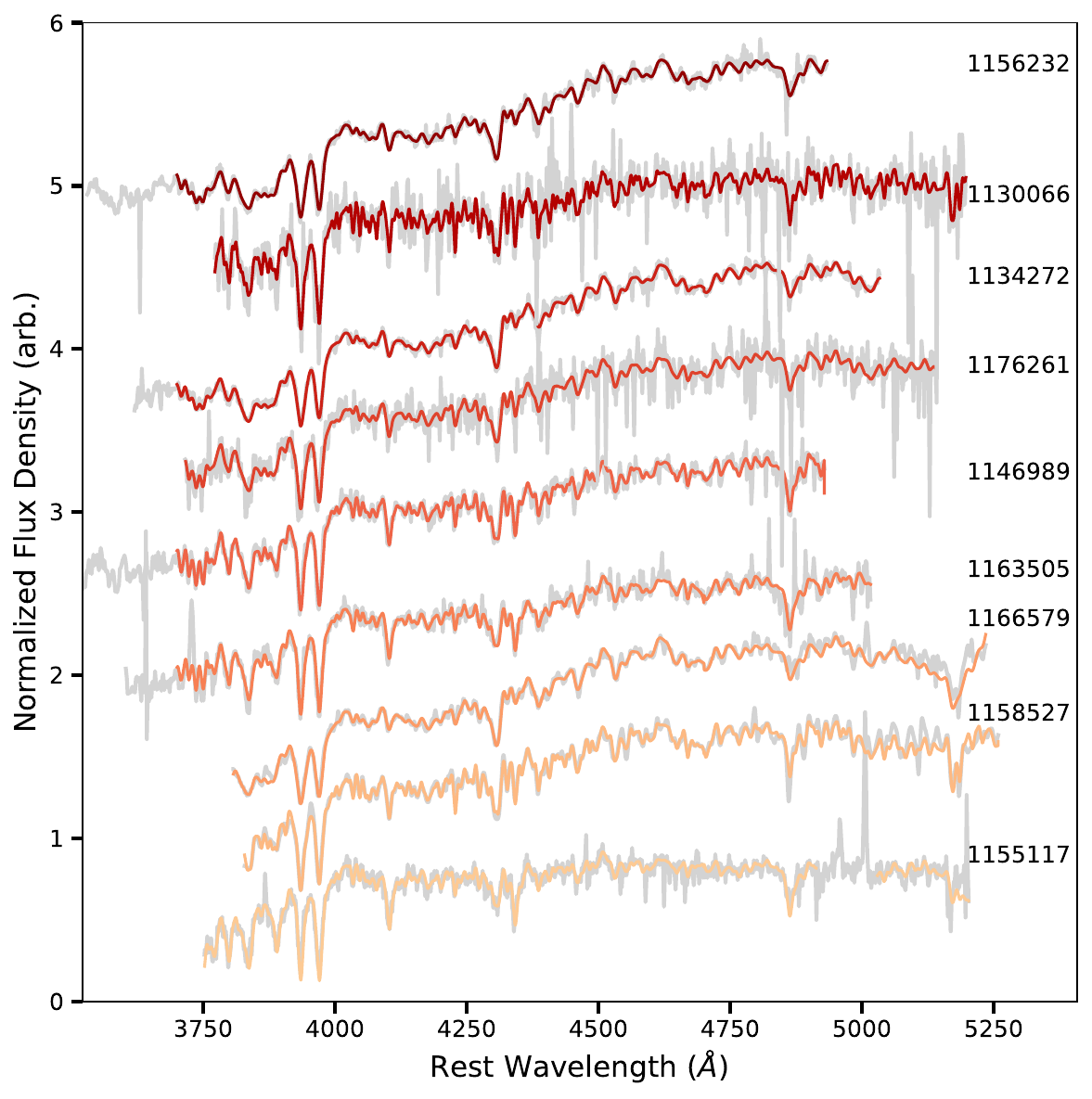}
    \caption{\textbf{Best fits to the bluer sides of the spectra from LEGA-C.}  Spectra are shown from bottom to top in order of increasing redshift.  The best-fitting models are over-plotted.  We normalize each spectrum and fit by the median value of the flux between $4,400$ and $4,500$ \AA.  We arbitrarily offset the spectra in the $y$ direction.  For the 6 spectra which we do not broaden, we median bin the data in bins of 5 pixels for visibility.}
    \label{fig:legac_fits}
\end{figure}

\begin{figure}
    \centering
    \includegraphics[width=\linewidth]{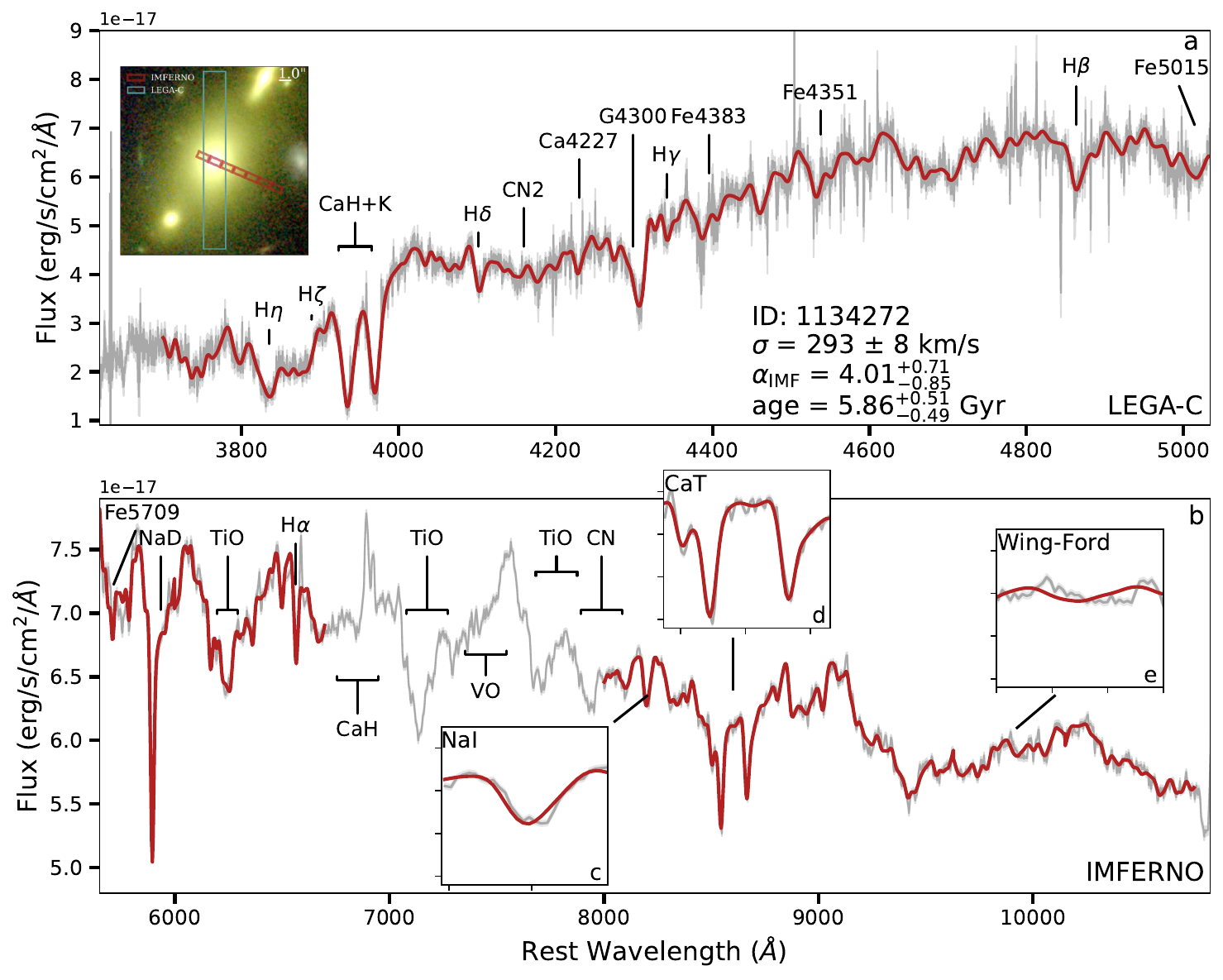}
    \caption{\textbf{Example \textsc{alf} fit to a bottom-heavy galaxy in our sample.}  The data are shown with shaded regions indicating one standard deviation on the flux, and the best-fit model over-plotted.  We show the spectrum from LEGA-C in \textbf{a} and the IMFERNO spectrum in \textbf{b}, but we note that we fit the entire wavelength range simultaneously.  In the inset panels (\textbf{c-e}), we zoom in on three strong absorption features which are sensitive to low-mass IMF variations.  We continuum-normalize the features in the inset panels, but include the continuum in the main panels.}
    \label{fig:fit_figure}
\end{figure}

\begin{figure}
    \centering
    \includegraphics[width=0.7\linewidth]{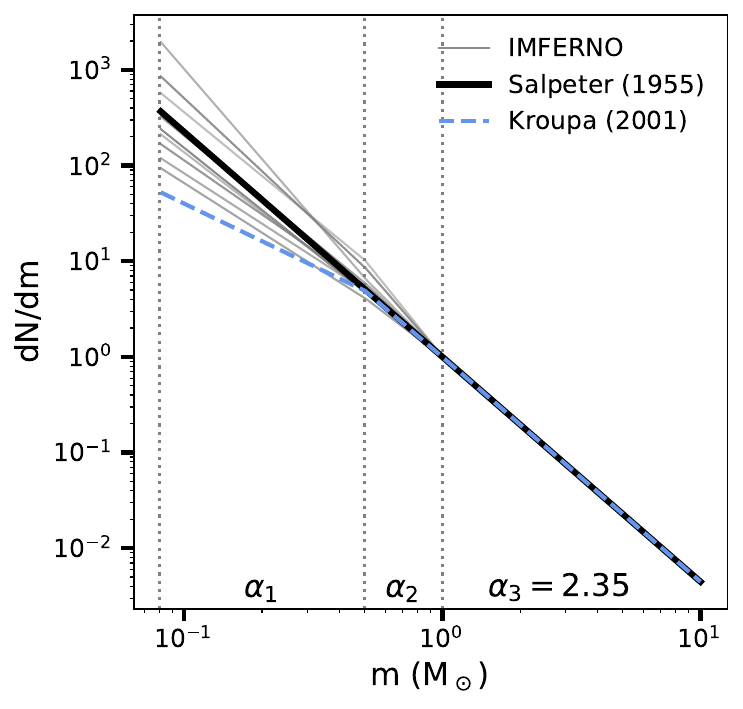}
    \caption{\textbf{IMFERNO IMF shapes compared to Kroupa and Salpeter.}  The IMF shapes for the IMFERNO sample are shown as thin, solid lines, and are compared to Kroupa (dashed line, \citep{Kroupa}) and Salpeter (thick, solid line, \citep{Salpeter}) IMFs.  The double broken power law IMF is parameterized as in equation~\ref{eq:2PL} and shown on the $y$ axis for a range of stellar masses ($x$ axis).  We indicate the breakpoints of the IMF with vertical dotted lines.}
    \label{fig:imf_slopes}
\end{figure}

\begin{figure}
\centering
\includegraphics[width=0.8\linewidth]{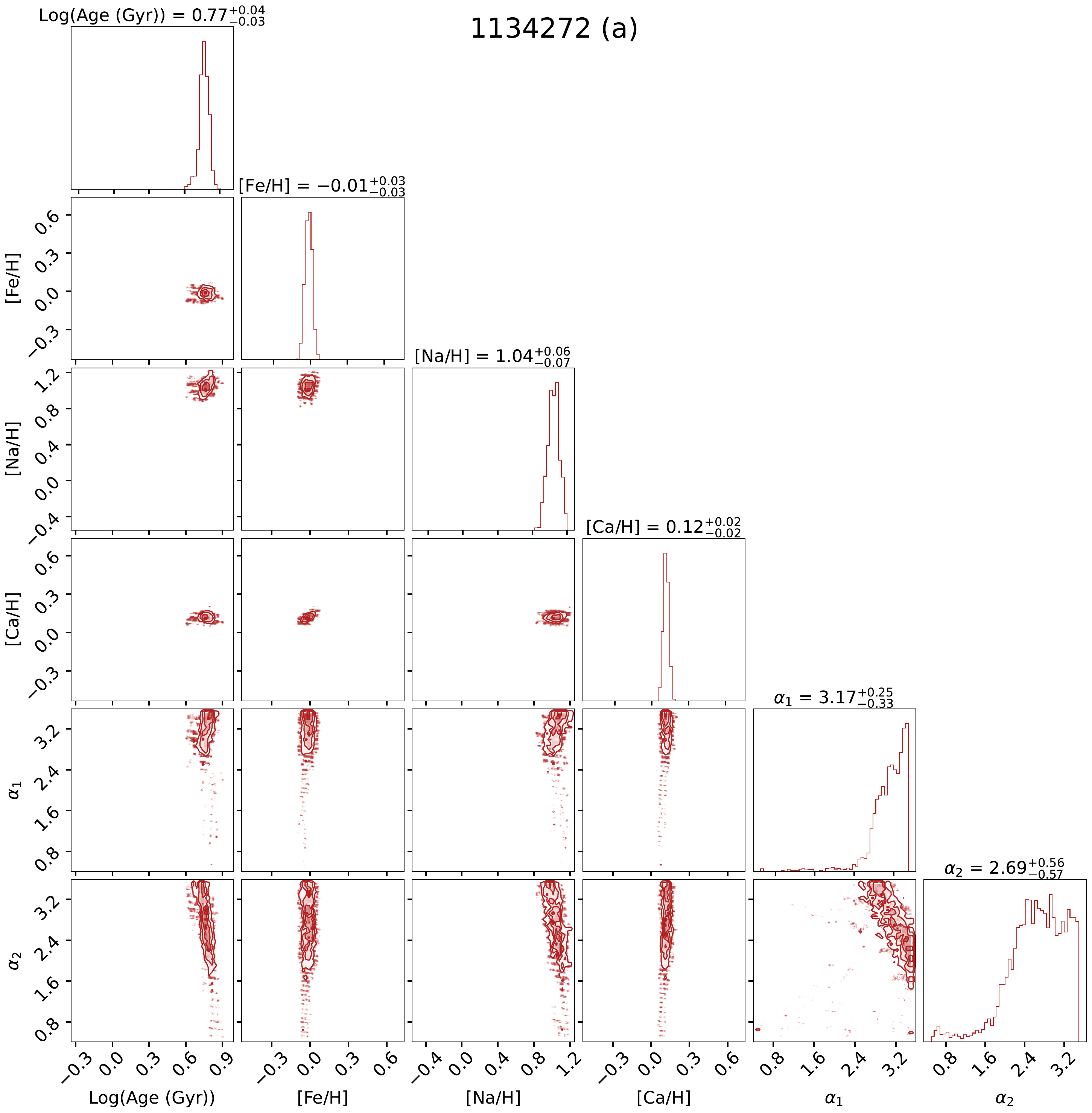}
\includegraphics[width=0.8\linewidth]{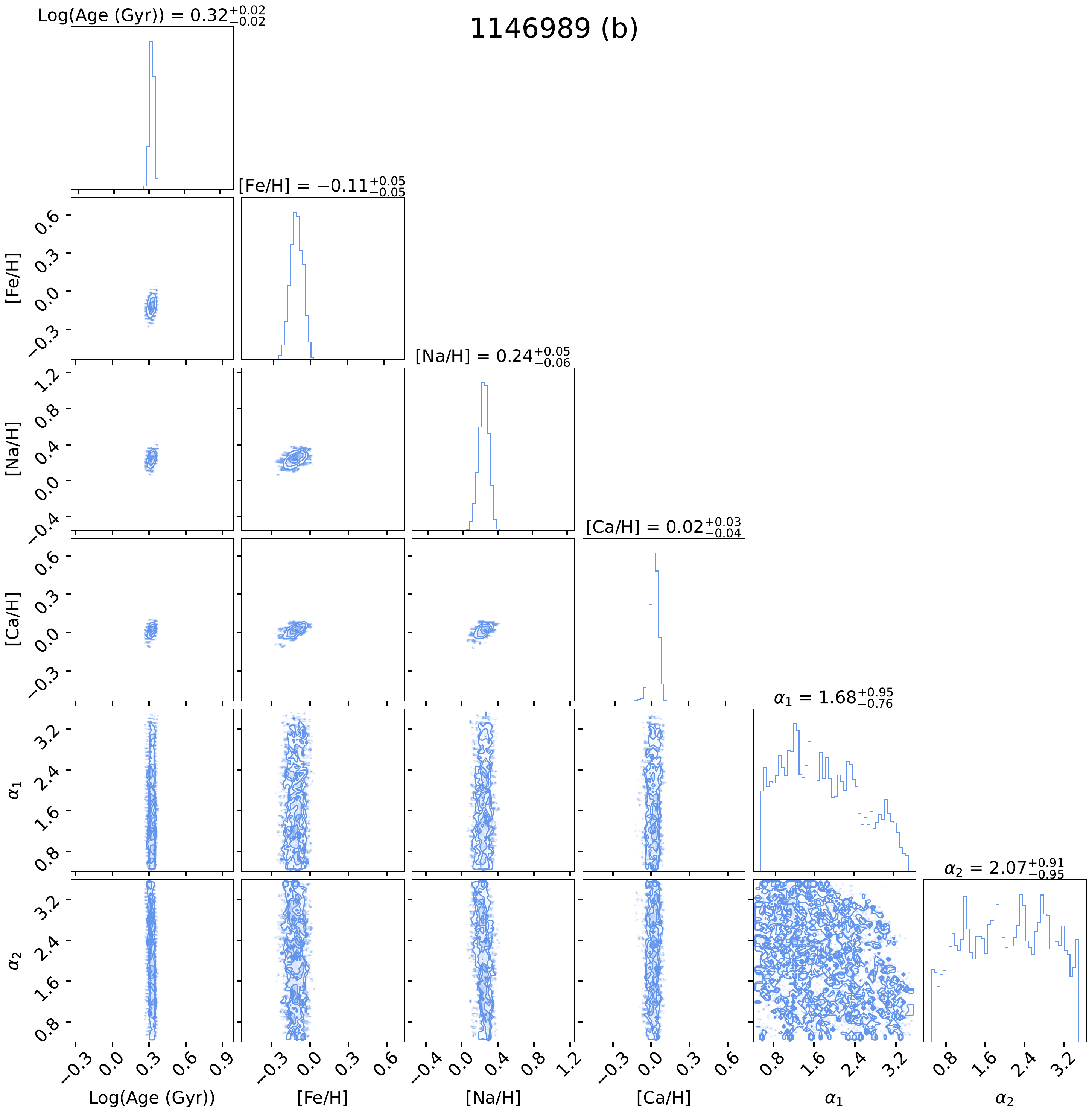}
\caption{\textbf{Corner plots for representative galaxies in our sample}.  For each galaxy, we show contours for $\log({\rm Age}{\ }{\rm (Gyr)})$, [Z/H], [Fe/H], [Na/H], [Ca/H], $\alpha_1$, and $\alpha_2$ (the two low-mass IMF slopes).  We show a bottom-heavy galaxy (1134272) in \textbf{a} and a Milky-Way-like galaxy (1146989) in \textbf{b}.}
\label{fig:corner_plots}
\end{figure}

\newpage
\FloatBarrier
\bmhead{Supplementary Information}
\FloatBarrier
\renewcommand{\figurename}{Supplementary Information Figure} 
\setcounter{figure}{0}
\begin{figure}
    \centering
    \includegraphics[width=\linewidth]{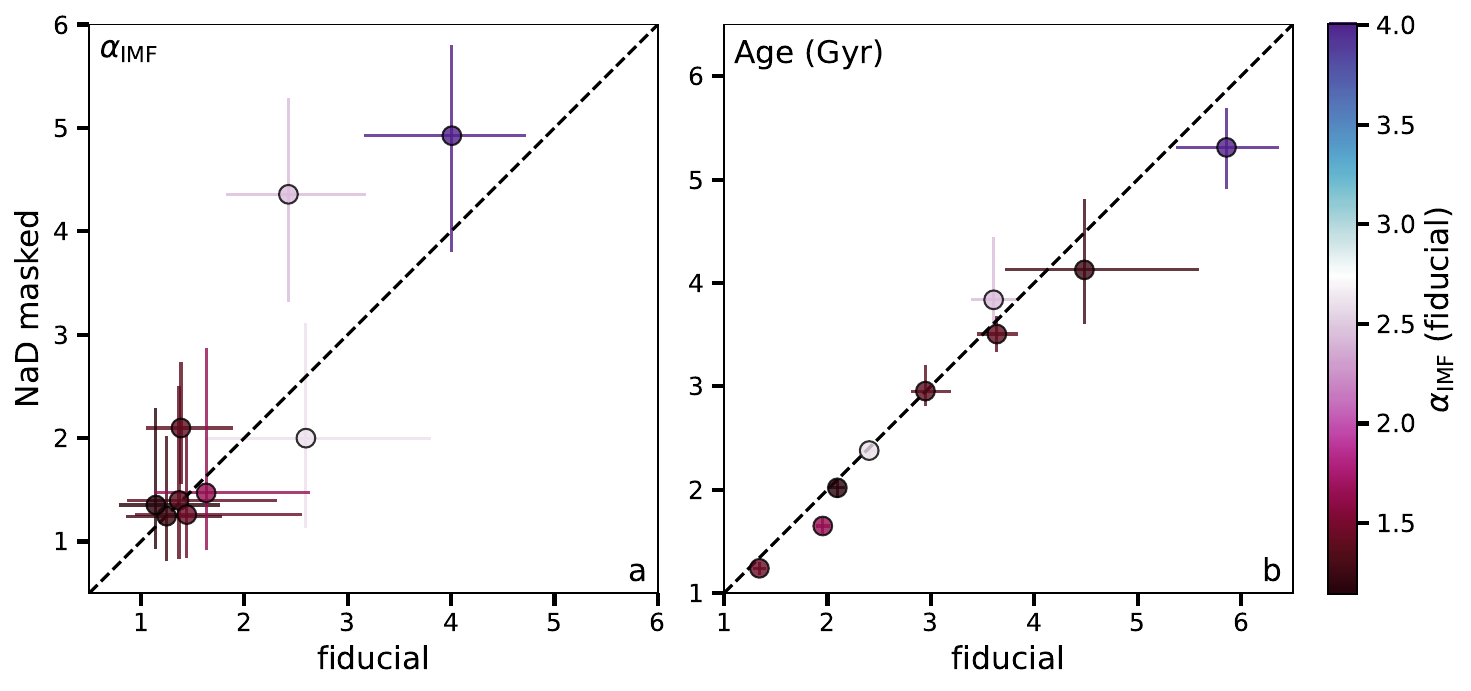}
    \caption{\textbf{The effect of masking the NaD feature for our sample.}  We compare $\alpha_{\rm IMF}$ (panel a) and age (panel b) from our fiducial fits and fits where we mask NaD.  We colour code each point by the $\alpha_{\rm IMF}$ that we obtain from our fiducial fits presented in the body of the paper.  Error bars represent statistical uncertainties at the level of 1 standard deviation.}
    \label{fig:NaD_effect}
\end{figure}

\begin{figure}
    \centering
    \includegraphics[width=\linewidth]{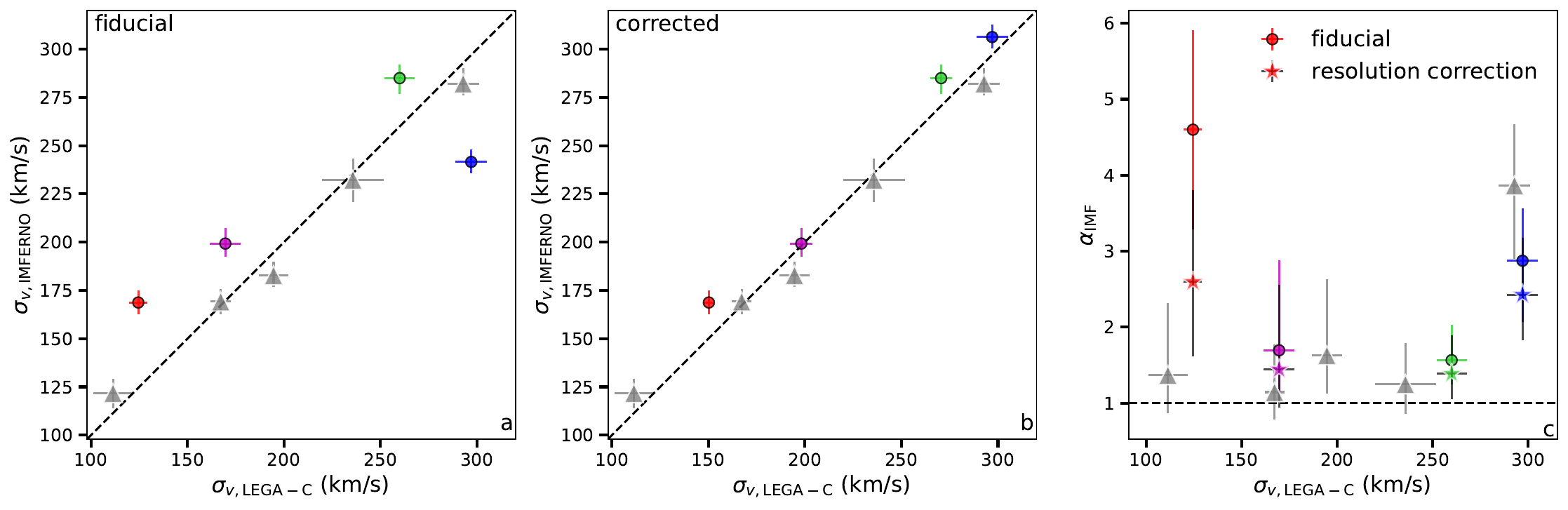}
    \caption{\textbf{Demonstration of our velocity dispersion correction to four galaxies in our sample.}  The triangles show the galaxies that do not need to be corrected, and all other points show the galaxies for which we perform the correction (i.e. broaden the side of the spectrum with lower $\sigma_v$).  Error bars represent statistical uncertainties at the level of 1 standard deviation.  In panel a, we show the fiducial $\sigma_v$ values from LEGA-C \citep{Bezanson_2018, van_der_Wel_2021} and from fits to the individual IMFERNO spectra.  In panel b, we show the corrected $\sigma_v$'s.  In panel c, we show the effect that the correction has on our derived $\alpha_{\rm IMF}$.}
    \label{fig:sigma_correction}
\end{figure}
\newpage

\bibliography{full_refs}

\end{document}